\documentclass[lettersize,journal]{IEEEtran}
\usepackage{amsmath,amsfonts}
\usepackage{algorithmic}
\usepackage{algorithm}
\usepackage{array}
\usepackage{amsmath}
\usepackage{amssymb}
\usepackage{mathtools}
\usepackage{booktabs}
\usepackage{stmaryrd}
\usepackage{enumitem}
\usepackage{textcomp}
\usepackage{pgfplots}
\pgfplotsset{compat=1.18}
\usepackage{tikz}
\usetikzlibrary{calc}
\usepackage{subcaption}
\usepackage{stfloats}
\usepackage{url}
\usepackage{verbatim}
\usepackage{graphicx}
\usepackage{colortbl}
\usepackage[hidelinks]{hyperref}
\usepackage{biblatex} %Imports biblatex package
\AtBeginBibliography{\footnotesize}

\usepackage{pdfpages}

\definecolor{cerulean}{rgb}{0.0, 0.48, 0.65}
\definecolor{darkgreen}{rgb}{0., 0.67, 0.}

\newcommand{\hg}{$H \ $}

\pgfmathdeclarefunction{decayFunction}{1}{%
    \pgfmathparse{exp(-1*(6-\x))}%
}

\newcommand{\convexpath}[2]{
	[   
	create hullnodes/.code={
		\global\edef\namelist{#1}
		\foreach [count=\counter] \nodename in \namelist {
			\global\edef\numberofnodes{\counter}
			\node at (\nodename) [draw=none,name=hullnode\counter] {};
		}
		\node at (hullnode\numberofnodes) [name=hullnode0,draw=none] {};
		\pgfmathtruncatemacro\lastnumber{\numberofnodes+1}
		\node at (hullnode1) [name=hullnode\lastnumber,draw=none] {};
	},
	create hullnodes
	]
	($(hullnode1)!#2!-90:(hullnode0)$)
	\foreach [
	evaluate=\currentnode as \previousnode using \currentnode-1,
	evaluate=\currentnode as \nextnode using \currentnode+1
	] \currentnode in {1,...,\numberofnodes} {
		-- ($(hullnode\currentnode)!#2!-90:(hullnode\previousnode)$)
		let \p1 = ($(hullnode\currentnode)!#2!-90:(hullnode\previousnode) - (hullnode\currentnode)$),
		\n1 = {atan2(\y1,\x1)},
		\p2 = ($(hullnode\currentnode)!#2!90:(hullnode\nextnode) - (hullnode\currentnode)$),
		\n2 = {atan2(\y2,\x2)},
		\n{delta} = {-Mod(\n1-\n2,360)}
		in 
		{arc [start angle=\n1, delta angle=\n{delta}, radius=#2]}
	}
	-- cycle
}

\begin{document}

\title{Modeling the Impact of Group Interactions on Climate-related Opinion Change in Reddit}

\author{Alessia Antelmi$^*$, Carmine Spagnuolo$^\dagger$, Luca Maria Aiello$^\ddagger$\\
$^*$Università degli Studi di Torino, Turin, Italy\\
$^\dagger$Università degli Studi di Salerno, Salerno, Italy\\
$^\ddagger$IT University of Copenhagen, Copenhagen, Denmark
\thanks{This work has been submitted to the IEEE for possible publication. Copyright may be transferred without notice, after which this version may no longer be accessible.}
}%end

\maketitle
% 100–250 word abstract 
\begin{abstract}
% === V2
Opinion dynamics models describe the evolution of behavioral changes within social networks and are essential for informing strategies aimed at fostering positive collective changes, such as climate action initiatives. When applied to social media interactions, these models typically represent social exchanges in a dyadic format to allow for a convenient encoding of interactions into a graph where edges represent the flow of information from one individual to another. However, this structural assumption fails to adequately reflect the nature of group discussions prevalent on many social media platforms. To address this limitation, we present a temporal hypergraph model that effectively captures the group dynamics inherent in conversational threads, and we apply it to discussions about climate change on Reddit. This model predicts temporal shifts in stance towards climate issues at the level of individual users. In contrast to traditional studies in opinion dynamics that typically rely on simulations or limited empirical validation, our approach is tested against a comprehensive ground truth estimated by a large language model at the level of individual user comments. Our findings demonstrate that using hypergraphs to model group interactions yields superior predictions of the microscopic dynamics of opinion formation, compared to state-of-the-art models based on dyadic interactions. Although our research contributes to the understanding of these complex social systems, significant challenges remain in capturing the nuances of how opinions are formed and evolve within online spaces.
\end{abstract}


%%%%%%%%%%%%%%%%%%%%%%%%%%%%%%%%%%%%%%%%%%%%%%%%%%%%%%%%%%%%%
% INTRODUCTION
%%%%%%%%%%%%%%%%%%%%%%%%%%%%%%%%%%%%%%%%%%%%%%%%%%%%%%%%%%%%%
\section{Introduction}\label{sec:intro}
% social media and opinions
% === V2
Online social media represent the largest arena for public discourse, where people gather news and form their opinions from discussions~\cite{bode16political,vaccari2021outside}. The vast amount of user-generated content on these platforms allows researchers to study and model how opinions and behaviors spread over complex social networks~\cite{das2014modeling}. Social opinion mining has, hence, emerged as a prominent discipline focused on ``examining user-generated content from individuals or groups who express their views on various entities, issues, events, or topics through social media interactions"~\cite{Cortis_AIR_2021}.
% importance of groups
In particular, the role \emph{groups} play has a critical role in opinion formation~\cite{Moussaid_PlosOne_2013}, and understanding people's group affiliations is fundamental to revealing personal behaviors~\cite{Forsyth2019}, as these are heavily influenced by the crowd~\cite{Edelson_Science_2011}. Group interactions are the building blocks of real-world social systems~\cite{Battiston_PhysicsReports_2020,Cencetti_SciRep_2021,March_1976}, and have been shown to be crucial in facilitating the spread of new opinions held by committed minorities~\cite{Iacopini_CommPhys_2022}, seeding and sustaining contagion processes~\cite{StOnge_CommPhys_2022}, and contributing to a rich phenomenology that includes abrupt transitions, bi-stability, and critical mass phenomena~\cite{AlvarezRodriguez_NatureHB_2021,Iacopini_CommPhys_2022,Iacopini_NatureComm_2019}.

\smallskip
% === V2
Conversational data extracted from social media represents a valuable source of information that can reveal patterns of social influence and the emergence of collective behaviors~\cite{Gonzalez_SN_2016,pera25extracting,Lucchini_RoyalSoc_2022}. Such digital traces enable researchers to map the complex interplay between individual decisions and group dynamics, offering unprecedented opportunities to test theoretical models against large-scale empirical evidence. 
Recent literature has primarily relied on simulations to describe the theoretical behaviors of opinion models that incorporate high-order interactions. However, to date, none of these studies have applied such models to real interaction data from social media. This gap arises from two significant challenges associated with quantifying group influence on opinion dynamics through social media data. The first challenge involves the design of appropriate models capable of capturing high-order interactions within user conversations. The second challenge pertains to the evaluation of various models of user interactions against ground-truth data, in order to assess their accuracy in replicating real-world processes. These critical issues shape our research questions (RQs).

\begin{itemize}[leftmargin=*]
    \item RQ1: How can we effectively capture high-order interactions in online social media conversations? How can we model the temporal evolution of the influence that comments can exert over time?

    \item RQ2: How do models incorporating high-order interactions compare to traditional approaches in representing real-world opinion diffusion processes? To what extent does incorporating high-order social dynamics improve the realism of such diffusion models?
\end{itemize}
%
%
% contribution summary
In this work, we addressed the above questions by:
\begin{itemize}[leftmargin=*]
    \item Designing a framework to model conversational networks via time-varying hypergraphs (TVHs). Hypergraphs are mathematical structures that generalize the concept of graphs by allowing each (hyper)edge to connect more than two nodes~\cite{Bretto2013}. This capability enables the abstraction of complex concepts such as group pressure and collective influence. Conversational TVHs capture the intuition that users may be more influenced by local parts of conversations rather than the whole thread and that old comments may not exert the same influence as the newer ones. We leveraged this novel framework to study opinion dynamics at scale.

    \item Developing an evaluation framework to assess the effectiveness of different network models in propagating opinions under the same experimental setting against ground-truth data. This process involved: \textit{(i)} building a novel stance-annotated dataset of climate change discussions to address the lack of ground-truth data, and \textit{(ii)} defining three metrics to quantitatively assess how accurately different network models capture real-world opinion dynamics.
 
\end{itemize}

% results
\noindent Our analysis revealed two main findings.
First, our results empirically demonstrate that hypergraph-based modeling of social dynamics offers advantages over traditional dyadic interaction approaches in reproducing opinion dynamics.  By modeling higher-order relationships, we can better represent how multiple users simultaneously influence each other within conversation threads, revealing the crucial role of collective interactions in opinion formation and evolution. This finding emphasizes that understanding group dynamics is essential when studying opinion formation in social networks, as these higher-order relationships appear to be fundamental drivers of the process.
Second, although our approach yields promising results, this research sheds light on a critical limitation in the field: the scarcity of ground-truth data about opinion dynamics. Such data is essential for capturing the complexity of opinion formation, which extends beyond observable online interactions, and hence, for informing the design, implementation, and evaluation of more robust models that can accurately represent real-world social processes.

\smallskip
\noindent \textbf{Reproducibility.}
The implementation and experimental code can be found at the following GitHub repository: \href{https://github.com/alessant/GrootSim}{github.com/alessant/GrootSim}. The datasets used in this study can be found at \href{https://zenodo.org/records/15225705}{zenodo.org/records/15225705}.

%%%%%%%%%%%%%%%%%%%%%%%%%%%%%%%%%%%%%%%%%%%%%%%%%%%%%%%%%%%%%
% BACKGROUND
%%%%%%%%%%%%%%%%%%%%%%%%%%%%%%%%%%%%%%%%%%%%%%%%%%%%%%%%%%%%%
\section{Background}\label{sec:background}
\subsection{Hypergraphs}
A \textbf{hypergraph} is an ordered pair $H=(\mathcal{V}, \mathcal{E})$, where $\mathcal{V}$ is the set of nodes, and $ \mathcal{E}$ is the set of hyperedges (Fig.~\ref{fig:hg}). Each hyperedge is a non-empty subset of nodes. The structure of a hypergraph is usually represented by an \textbf{incidence matrix} $\mathbf{H} \in \{0,1\}^{|\mathcal{V}| \times |\mathcal{E}|}$, with each entry $\mathbf{H}(v,e)$ indicating whether the vertex $v$ is in the hyperedge $e$, i.e., \mbox{$\mathbf{H}(v,e) = \llbracket v \in e \rrbracket$.}

\smallskip

The \textbf{clique graph} of \hg is the graph, denoted with $[H]_2$, whose vertices are the vertices of \hg and where two distinct vertices form an edge if and only if they are in the same hyperedge of $H$. %(see Figure~\ref{fig:hg_clique}). 
In other words, each hyperedge of \hg appears as a complete sub-graph in $[H]_2$.

\subsection{A non-linear consensus process on hypergraphs}\label{subsec:diffusion_model}
The literature on opinion formation and dynamics in online social media is extensive~\cite{Noorazar_2020}. Recently, there has been growing interest in incorporating group dynamics into these processes, although this remains a highly unexplored field~\cite{Battiston_PhysicsReports_2020}. 
In this work, we used the multi-body consensus model (MCM) proposed by Sahasrabuddhe et al.~\cite{Sahasrabuddhe_2021} for the following reasons: \textit{(i)} it combines homophily and group pressure/conformity social theories, extending them to hypergraphs; \textit{(ii)} it can be applied to both graphs and hypergraphs; and \textit{(iii)} by adjusting user involvement parameters, we can simulate various outcomes, including polarization, consensus toward the mean or a combination of both -- allowing us to explore diverse scenarios.
Formally, MCM can be described by two simplified models as follows:

\begin{figure}[t!]
    \centering
    \resizebox{.6\linewidth}{!}{%
    \begin{tikzpicture}[rotate=270]
        \node (a) at (1., 5.6) {}; %v1
        \node (b) at (2, 3.6) {}; %v2
        \node (c) at (0., 3.6) {}; %v3
        \node (d) at (1., .5) {}; %v4
        \node (e) at (1., 8.) {}; %v5
        \node (f) at (2., 6.9) {}; %v6
        \node (g) at (0., 6.9) {}; %v7
    
        \fill (a) circle (0.05) node [right] {\Large \textbf{$v_1$}};
        \fill (b) circle (0.05) node [left] {\Large \textbf{$v_2$}};
        \fill (c) circle (0.05) node [left] {\Large \textbf{$v_3$}};
        \fill (d) circle (0.05) node [right] {\Large \textbf{$v_4$}};
        \fill (e) circle (0.05) node [right] {\Large \textbf{$v_5$}};
        \fill (f) circle (0.05) node [right] {\Large \textbf{$v_6$}};
        \fill (g) circle (0.05) node [right] {\Large \textbf{$v_7$}};
        
        \draw[thick] \convexpath{c,a,b}{0.77cm};
        \draw[thick] \convexpath{c,d}{0.6cm};
        \draw[thick] \convexpath{b,d}{0.7cm};
        \draw[thick] \convexpath{a,g,e,f}{0.75cm};
        
        \begin{scope}[every node/.style = {node distance = 20pt}]
            \node[fill=white, rectangle] (e1) at (-0.2, 4.9) {\Large $e_1$};
            \node[fill=white, rectangle] (e2) at (0., 2.) {\Large $e_2$};
            \node[fill=white, rectangle] (e3) at (2.2, 2.) {\Large $e_3$};
            \node[fill=white, rectangle] (e4) at (2.6, 6.5) {\Large $e_4$};
        \end{scope}
    \end{tikzpicture}
}
    \vspace*{-0.7cm}
    \caption{An example of a hypergraph with four hyperedges.}
    \label{fig:hg}
\end{figure}
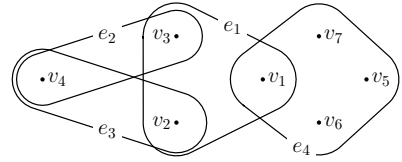

% MCM1
\begin{equation}
    \scriptstyle
    MCM_{I} \equiv \ {x_v}^{e_i} = 
    s_v^I\left(\left|\frac{\sum_{u \in e_i} x_u}{\left|e_i\right|}-x_v\right|\right) 
    \left( {\sum_{u \in e_i} (x_u - x_v)} \right)
\end{equation}

\begin{equation}
    \scriptstyle
    MCM_{II} \equiv \ {x_v}^{e_i} = 
    \sum_{u \in e_i} s_v^{II}\left(\left|\frac{\sum_{z \in e_i, z \neq v} x_z}{\left|e_i\right|-1}-x_u\right|\right)\left(x_u-x_v\right),
\end{equation}
whose combination $MCM_{I} \times MCM_{II}$ gives the final effect of a hyperedge $e_i$, with $e \in \mathcal{E}$ and $i \in \{1,2, \dots, |\mathcal{E}| \}$, on the state of a node $v \in e_i$.
The sociological motivation and mathematical properties of the two models are captured by the two modulating functions $s_v^{I}(\cdot)$ and $s_v^{II}(\cdot)$. In detail:
\begin{itemize}[leftmargin=*]
    % s1 homophily
    \item $s_v^{I}(\cdot)$ models the concept of homophily, which captures the tendency of like-minded individuals to interact~\cite{McPherson_AnnuRev_2001}. It is a function of the distance of $x_v$ to the mean state of the hyperedge $e_i$, and determines the rate at which $e_i$ influences the state of $v$. %In other words, $MCM_{I}$ modulates the competing effect of different hyperedges on the state of an incident node.
    The argument of the function quantifies the difference between the opinion of an individual $v$ and the average opinion of the group $e_i$ that $v$ belongs to. 

    \smallskip

    % s2 group pressure
    \item $s_v^{II}(\cdot)$ models the concept of conformity, which is used to describe the tendency of an individual to align their beliefs to those of their peers and is usually affected by the reinforcing nature of shared opinions (peer pressure)~\cite{Sherif_1961}. It is a function of the distance of a participating node $u$ from the mean state of the hyperedge $e_i$ excluding~$v$. The argument of the function quantifies the difference between the opinion of an individual $v$ to the average opinion of the group $e_i$ except~$v$.
\end{itemize}

\smallskip
\noindent In our experiments, we used the same definition for $s_v^{I}(\cdot)$ and $s_v^{II}(\cdot)$ as proposed by Sahasrabuddhe et al.~\cite{Sahasrabuddhe_2021}, which draw inspiration from the Social Judgement Theory~\cite{Sherif_1961} and the Jager–Amblard model~\cite{Jager_CMOT_2005}. In more detail, the homophily function $s_v^{I}$ is defined as:
$$
    %\scriptstyle
    s_v^I(x)= 
        \begin{cases}
            %\scriptstyle
            \mathrm{e}^{\lambda_v x} & x \leqslant \phi_v^A \\
            %\scriptstyle
            0 & \phi_v^A<x<\phi_v^R . \\ 
            %\scriptstyle
            -\mathrm{e}^{\lambda_v x} & x \geqslant \phi_v^R
        \end{cases}
$$
where $\phi_v^A$ and $\phi_v^R$ represent the threshold of acceptance or rejection of $v$ regarding an opinion, respectively, and $\lambda_v$ the stubbornness of each node. 
The group pressure function $s_v^{II}$ is described by an exponential modulating function $$s_v^{II}(x)=\mathrm{e}^{\delta_v x},$$ where $\delta_v$ represents the tendency of being influenced by like-minded individuals. %, with $\delta_v=-5$~\cite{Sahasrabuddhe_2021}.

%%%%%%%%%%%%%%%%%%%%%%%%%%%%%%%%%%%%%%%%%%%%%%%%%%%%%%%%%%%%%
% DATA AND METHODOLOGY or materials and methods
%%%%%%%%%%%%%%%%%%%%%%%%%%%%%%%%%%%%%%%%%%%%%%%%%%%%%%%%%%%%%
%\newpage
\section{Data \& Methodology}\label{sec:data_methodology}

\subsection{A climate-related Reddit dataset}
% A brief overview of Reddit
Reddit is one of the most popular social network websites focused on news aggregation and discussion, mainly in the form of question and answer. The key element of this platform is its organization in subreddits, which represent online communities centered around a specific topic, like technology, politics, or sports. 

%
% crawling process
% filtering process
% - giant component
% - no bots
% - no one-timers as leaves
% - all submissions with no one-time commenters
Since our study aimed to examine potential shifts in public opinion regarding climate change, we considered the California wildfires of 2021 and 2022 as a specific exogenous event that sparked significant discussion on the platform. We collected public data from Reddit using the Pushift API~\cite{Baumgartner_ICWSM_2020}. Our data collection process began in April 2023 by retrieving recent comments containing the keywords ``California" and ``wildfire".  This led us to identify the six most active subreddits on this topic: r/California, r/AskReddit, r/politics, r/news, r/collapse, and r/bayarea. We retrieved all submissions and comments in these subreddits spanning from July 1, 2020, to December 31, 2022. To isolate climate change-related content, we applied keyword filtering using terms such as ``climate action," ``climate denial," and ``global warming". We refined our dataset by excluding bots and one-time contributors who received no replies. Table~\ref{tab:dataset_stats} summarizes the main statistics of our dataset after preprocessing it.

% number of unique users, submissions, and comments
\begin{table}[b!]
    \centering

    \caption{Number of users, submissions, and comments in our dataset.}
    \label{tab:dataset_stats}

    \begin{tabular}{lrrr}
        \toprule
        
        \multicolumn{1}{c}{\textbf{Subreddit}} & 
        \multicolumn{1}{c}{\textbf{Unique users}} & 
        \multicolumn{1}{c}{\textbf{Submissions}} & 
        \multicolumn{1}{c}{\textbf{Comments}} \\

        \midrule
        
         r/AskReddit & 
         10,953 & 
         2,618 & 
         49,329 \\

        \rowcolor{gray!10}
        r/bayarea & 
        1,895 & 
        181 & 
        7,743 \\

        r/California & 
        1,743 &
        148 & 
        7,353 \\

        \rowcolor{gray!10}
        r/collapse & 
        21,532 & 
        2,429 & 
        194,695 \\
         
        r/news & 
        14,159 & 
        385 & 
        60,509 \\

        \rowcolor{gray!10}
        r/politics & 
        10,407 & 
        490 &
        43,721 \\

        \textbf{Total} & 
        54,923 & 
        6,251 & 
        363,350 \\

        \bottomrule
    \end{tabular}
\end{table}

\subsection{Modeling conversational networks with hypergraphs}
\label{subsec:hg_conversational_nets}
\begin{figure}[t!]
    \centering
    \includegraphics[width=.9\columnwidth]{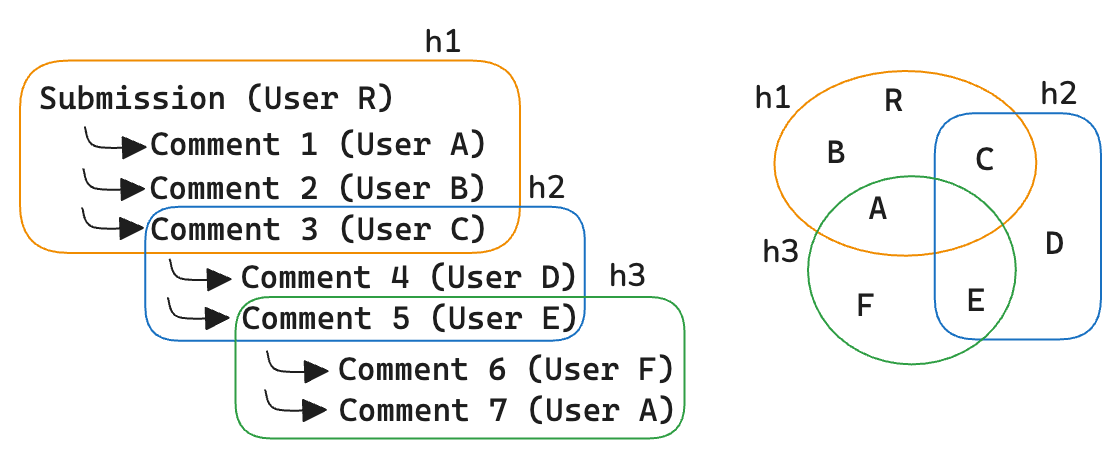}
    \caption{A toy example of a Reddit discussion (left) and its representation as a hypergraph (right).}
    \label{fig:conv_hypernet}
\end{figure}

To capture the high-order interactions in conversational threads and study opinion spread in online discussions, we employed hypergraphs. In this model, each user corresponds to a node, while each hyperedge represents a subthread within a submission (i.e., a conversation) where users interact. In other words, a hyperedge includes all users who have discussed in the same \textit{local} conversation.
Fig.~\ref{fig:conv_hypernet} illustrates this model.

\smallskip

\begin{figure}[b!]
    \centering

    \resizebox{0.9\linewidth}{!}{%
    \begin{tikzpicture}
    
        % % axes
        % \draw[->, thick] (-1,0)--(5,0) node[right]{$x$};
        % \draw[->, thick] (0,-1)--(0,3.5) node[above]{$y$};

        % %lines
        % %\draw[help lines, color=gray!30, dashed] (-4.9,-4.9) grid (4.9,4.9);
        % \draw[-, dashed] (1,0)--(1,0);

        % Coordinate system
        \draw[->] (-1,0) -- (7.6,0) node[right] {};
        \draw[->] (-0.5,-0.6) -- (-0.5,5) node[above] {};

        % axes labels
        \node at (3.5, -1) {\small Event timeline};
        \node[rotate=90] at (-0.8, 2.4) {\small Event importance};
        
        % Horizontal lines (light gray grid)
        % \foreach \y in {-1, 0, 1, 2, 3}
        %     \draw[gray!30] (-1,\y) -- (7,\y);

        % Purple dashed horizontal line
        \draw[purple, dashed, thick] (-0.5, 0.5) -- (7.53, 0.5);
        
        % Small tau symbol near purple line
        \node[purple] at (6.7, 0.7) {$\tau$};
        
        % Green curve (decay function)
        \draw[
            color=green!70!black, 
            thick,
            domain=0:6
        ] plot (\x, {exp(-1*(4.63-\x))}) node {};

        \node[green!70!black] at (5., 2.) {$f$};
        
        % \begin{axis}[
        %     %unbounded coords=discard,
        %     axis line style={opacity=0},
        %     x tick label style={opacity=0},
        %     y tick label style={opacity=0},
        %     ticks=none
        % ]
        %     \addplot[color=green!70!black, thick, domain=0:6] (\x,{decayFunction(\x)});
        % \end{axis}

        %\draw[green!70!black, thick] (1,0.2) .. controls (3,0.5) and (4,1.5) .. (6,3.5);
        
        % Vertical marks along x-axis
        \draw[thick] (1,0) -- (1,-0.2);
        \draw[thick] (1.7,0) -- (1.7,-0.2);
        \draw[thick] (2.5,0) -- (2.5,-0.2);
        \draw[thick] (3.3,0) -- (3.3,-0.2);
        \draw[thick] (4.3,0) -- (4.3,-0.2);
        \draw[thick] (5.1,0) -- (5.1,-0.2);
        \draw[thick] (6,0) -- (6,-0.2);
        
        % Labels below x-axis
        \node at (1,-0.5) {$d_i$};
        \node at (1.7,-0.5) {$d_{i+1}$};
        \node at (2.5,-0.5) {$d_{i+2}$};
        \node at (3.3,-0.5) {$d_{i+3}$};
        \node at (4.3,-0.5) {$d_{i+4}$};
        \node at (5.1,-0.5) {$d_{i+5}$};
        \node at (6,-0.5) {$d_{i+6}$};

        \draw[dotted, thick] (0., -0.5) -- (0.5, -0.5);
        \draw[dotted, thick] (6.6, -0.5) -- (7.1, -0.5);

        % Yellow/gold horizontal line with slight tilt
        \draw[yellow!70!brown, thick] (3.9, 0) -- (6, 0);
        
        % Vertical bars/divisions
        \draw[gray, thick] (6,0) -- (6,4);
        \draw[gray, dashed, thick] (3.9,0) -- (3.9,4);
        
        % Downward-pointing arrow near the curve
        %\draw[->, thick] (5,2.8) -- (5,2);
        
        % Triangle symbols
        \node[cerulean] at (3.2, 3.4) {\footnotesize $\Delta$};
        \node[cerulean] at (6.7, 3.4) {\footnotesize $\Delta$};

        % draw ∆ windows
        \draw[|-|, cerulean] (0.05, 3.2) -- (3.49, 3.2);
        \draw[|-|, cerulean] (3.5, 3.2) -- (7, 3.2);

        % before - after
        \draw[cerulean, thick, dotted] (-0.5, 3.2) -- (0.03, 3.2);
        \draw[cerulean, thick, dotted] (7, 3.2) -- (7.53, 3.2);

        \draw[cerulean, thick, dotted] (3.5, 3.1) -- (3.5, 0);

        % draw s∆ windows
        % first line
        \draw[gray, ->] (0.05, 3.6) -- (3.5, 3.6);
        \draw[gray, dotted] (-0.5, 3.6) -- (0.05, 3.6);
        \node[gray] at (0.7, 3.8) {\scriptsize $(s-1) \cdot \Delta$};
        
        \draw[gray, ->] (0.05, 4.1) -- (7, 4.1);
        \draw[gray, dotted] (-0.5, 4.1) -- (0.05, 4.1);
        \node[gray] at (0.4, 4.3) {\scriptsize $s \cdot \Delta$};

        \draw[gray, -] (0.05, 4.6) -- (7, 4.6);
        \draw[gray, dotted] (-0.5, 4.6) -- (0.05, 4.6);
        \draw[gray, dotted, ->] (7, 4.6) -- (7.53, 4.6);
        \node[gray] at (0.7, 4.8) {\scriptsize $(s+1) \cdot \Delta$};

        % Curve connecting the triangles
        %\draw[] (0, 0) .. controls (1.75, -1.5) .. (3.5, 0);
        %\draw[thick] (0, 0) .. controls (2,-1) and (4,-2) .. (5,-1.2);

    \end{tikzpicture}
}

    \vspace{-0.7cm} 
    
    \caption{The construction process of a time-varying conversational hypergraph. In this example, $D_{s\Delta} = [d_1,..., d_i,..., d_{i+6}] $, while $\widetilde{ \mathcal{D}}_{s\Delta} = \{d_{i+4}, d_{i+5}, d_{i+6} \}$. 
    }
    \label{fig:conv_TVH}

\end{figure}
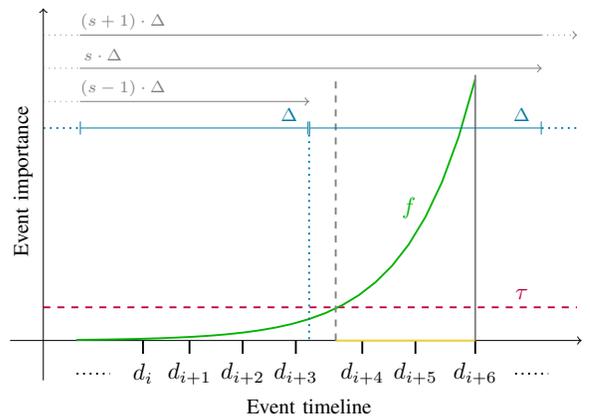

Based on the assumption that old comments may not exert the same influence as the newer ones, we incorporated a temporal dimension into this framework. The temporal axis is regulated by three parameters: \textit{(i)} a stride $\Delta$ that controls how many new events (or time intervals) we consider at each new iteration; \textit{(ii)} a decay function $f$ describing how quickly past events (e.g., comments) lose importance; and \textit{(iii)} an event importance threshold $\tau$, that represents the minimum weight an event must have to influence future comments (see Fig.~\ref{fig:conv_TVH}). In this work, we use an exponential decay function $f=e^{-t}$ to weight events based on their temporal distance $t$ from the most recent post in the analysis period, as such functions represent a common choice to regulate the importance of data over time in recommender systems~\cite{kille_RecSysWS_2017,Bogina_UMUAI_2023}. 
In the following, we formally define the concepts introduced. %of such conversational time-varying hypergraphs.

\smallskip

Let us describe a set of conversations $D$ as an ordered sequence of tuples $d_i=(u_{i}, v_{i}, c_{i}, t_{i})$, indicating that a user $u$ replied to the user $v$ with some text $c$ at a given time~$t$. Hence, $D \coloneq \langle d_i \rangle_{i=1}^N$.
Given a time step $s \in \mathbb{N}^+$ and a temporal window $\Delta$, we define a subsequence of conversations up to the timestamp $s \cdot \Delta$ as $\mathcal{D}_{s\Delta} \coloneq \{ d_i=(u_{i}, v_{i}, c_{i}, t_{i}) \ | \ t_{i} < s \cdot \Delta \}$.

\smallskip

Then, let us introduce a filtering function $\phi$ that, given a decay function $f: \mathbb{N} \rightarrow \mathbb{R}$ and a weight threshold $\tau \in \mathbb{R}^+_0$, filters out all content whose importance is too low to further influence the conversation. Formally, 
\begin{center}
    $\widetilde{ \mathcal{D}}_{s\Delta} =
    \phi(f, \tau, \mathcal{D}_{s\Delta}) \coloneq \{ d_i=(u_{i}, v_{i}, c_{i}, t_{i}) \ | \ f(t_{i}) \geq \tau \}$. 
\end{center}

\noindent
In other words, at each time step $s$, we advance by a stride~$\Delta$, and select all events $d_i$ whose score $f(t_i)$ exceeds the threshold $\tau$. This approach enables three possible configurations. We can include \textit{(i)} all events within the current observation window $\Delta$, \textit{(ii)} a subset of those, or \textit{(iii)} events from previous observation window(s) in addition to current events (considering the interval $s \cdot \Delta$). See Fig.~\ref{fig:conv_TVH} for an example. Formally, based on the above definitions, at each new time step, we consider all events
$
    \{ d_i \ | \ t_i < s \cdot \Delta \ \land \ f(t_i) \ge \tau \}.
$

\smallskip

Let $H_s = (\mathcal{V}_s, \mathcal{E}_s)$ be the conversational hypergraph built on top of the filtered dataset $\widetilde{ \mathcal{D}}_{s\Delta}$, where 
\begin{center}
    $
        \mathcal{V}_s \ \equiv \ \{ u \ | \ (u, \cdot, \cdot, \cdot) \in \widetilde{ \mathcal{D}}_{s\Delta} \} \ \text{and}
    $
\end{center}
\begin{center}
    $
    \mathcal{E}_s \ \equiv \  \{ 
        \{ u \ | \ \exists (u, v, \cdot, \cdot) \} \cup \{ v \}
    \}
    $.
\end{center}

\noindent The union of all $H_s \ \forall s \in S $ can be seen a conversational time-varying hypergraph~\cite{Antelmi_AAMAS_2020} $H=(\mathcal{V}, \mathcal{E}, \mathcal{T}, \rho)$, where 
\begin{center}
    $\mathcal{V} \equiv \bigcup_{s \in S} \mathcal{V}_{H_{s}} $,
\end{center}
\begin{center}
    $ \mathcal{E} \equiv \bigcup_{s \in S} \mathcal{E}_{H_{s}}$,
\end{center}
$\mathcal{T}$ is the lifetime of the system, with
\begin{center}
     $\mathcal{T} \equiv [t_{0}, t_{max}]$, 
\end{center}
and $\rho$ is an indicator function, with 
\begin{center}
    $\rho(e, s) = 1 \ \text{if} \ e \in H_s, 0 \ \text{otherwise}$.
\end{center}

\subsection{Model evaluation}
To address our RQs, we needed to define two key elements in our analysis pipeline: the labeling process to evaluate each user's stance on climate change and the metrics used to assess the performance of each network model.

\smallskip

\subsubsection{Ground-truth construction}
To assign labels representing each user's stance on climate change for every piece of text in our dataset, we used the GPT-3.5 Turbo model. We approached this task as a stance detection problem, drawing inspiration from the methodology used to create the SPINOS dataset~\cite{Sakketou_ACL_2022}. This dataset was designed to capture subtle opinion shifts and detect fine-grained stances over time. It includes detailed annotations of stance polarity and intensity per user, both in long-term and short-term conversational threads, enabling the detection of nuanced opinion changes. The dataset was primarily annotated by non-experts, with a portion of the data also reviewed by experts.

Our labeling prompt\footnote{See Section 1 in Supplemental Material.} was adapted from the stance detection task given to the annotators to build the SPINOS dataset. We validated this approach using SPINOS, and while the GPT-4 Turbo model yielded better performance, we opted to use GPT-3.5 due to its reasonable accuracy at a lower cost. Specifically, GPT-3.5 correctly identified the stance of positive/negative comments in 90\% of the cases. This decision enabled us to label our large-scale dataset efficiently without incurring prohibitive expenses.
The model assigned output labels on a scale from 0 to 1 to each comment, where a score of 0.0 stands for statements expressing views \textit{strongly against}, 0.5 for those taking a \textit{neutral} position, and 1.0 for statements \textit{strongly in favor} of the stance ``Climate change is a real concern.".

\smallskip
\subsubsection{Evaluation metrics}\label{subsubsec:eval_metrics}
We defined three metrics to evaluate and quantify how well a network model conveys information about real-world opinion dynamics. 
The first two evaluation metrics rely on the key observation that the first time a user changes their opinion marks a turning point in their belief system, as empirical evidence shows that users tend to modify the intensity of their stance gradually rather than suddenly reversing their position entirely~\cite{Moussaid_PlosOne_2013,Sakketou_ACL_2022}. Specifically, the first criterion (\textbf{\underline{C1}}) measures the capacity of the model to identify a user's first opinion drift (occurring when two consecutive opinions belong to different classes). The second (\textbf{\underline{C2}}) measures the average number of intervals by which the model anticipates or postpones the identification of a user's first opinion drift, evaluated using the root mean square error (RMSE) on true positive instances, i.e., users for whom the model correctly identified their first opinion change. 
The third evaluation metric serves as a performance indicator as it directly measures the model's effectiveness in simulating real-world opinion evolution patterns over the complete observation period. In particular, this last criterion (\textbf{\underline{C3}}) assesses how well each model predicts users' final opinions. 
To evaluate each network model for criteria C1 and C3, we treat the problem similarly to a (multi-class) classification task by discretizing the continuous opinion vector into distinct categories (see Section~\ref{subsubsec:eval} for details). %and employ standard evaluation metrics, e.g., accuracy.

%%%%%%%%%%%%%%%%%%%%%%%%%%%%%%%%%%%%%%%%%%%%%%%%%%%%%%%%%%%%%
% RESULTS
%%%%%%%%%%%%%%%%%%%%%%%%%%%%%%%%%%%%%%%%%%%%%%%%%%%%%%%%%%%%%
\section{Results}\label{sec:results}

\subsection{Experimental setting}\label{subsec:exp_setting}
We compared our hypergraph-based model and traditional graph-based approaches to evaluate whether incorporating high-order social dynamics into diffusion models better reflects real-world opinion change processes. Specifically, our comparison framework included \textit{(i)} the proposed hypergraph-based conversational model $H$ discussed in Section~\ref{subsec:hg_conversational_nets}; \textit{(ii)} the corresponding clique-based representation $[H]_2$ of the hypergraph $H$; and \textit{(iii)} the conventional reply-based graph model $G$, where an edge connects two users u and v if and only if v replied to u. We treated interactions as undirected for all three models, as the opinion diffusion model considered in this study does not account for directionality.

\smallskip
\subsubsection{Network model parameters}
We applied the same set of parameters for all three network models. We fixed a time window $\Delta=1$ day and a minimum importance threshold~$\tau$ of $0.1$ to filter relevant events. 

\smallskip
\subsubsection{Diffusion model parameters} 
To capture the empirical distribution of users' final opinions characterized by partial consensus and emerging polarization, we configured the model parameters according to the original authors' specifications, setting $\phi_v^A = 0.15$ for the acceptance threshold, $\phi_v^R = 0.30$ for the rejection threshold, $\lambda_v=-1$, and $\delta_v=-5$~\cite{Sahasrabuddhe_2021}.

\smallskip
\subsubsection{Data} 
% **Dataset(s)**
% 1. All six subreddits together
% 2. Single subreddits
%     1. AskReddit
%     2. bayarea
%     3. California
%     4. collapse
%     5. news
%     6. politics
% 3. Subreddits pairs
%     1. collapse, news
%     2. collapse, politics
%     3. askreddit, news
%     4. askreddit, collapse
%     5. askreddit, politics
%     6. bayarea, news
% 4. Triples of subreddits
%     1. collapse, news, politics
%     2. AskReddit, news, politics
%
We conducted the same set of experiments on the conversational (hyper)networks built from the following sources: \textit{(i)} each of the six individual subreddits (see Table~\ref{tab:dataset_stats}); \textit{(ii)} specific pairs of subreddits: (collapse, news), (collapse, politics), (askreddit, news), (askreddit, collapse), (askreddit, politics), and (bayarea, news); \textit{(iii)} two combinations of three subreddits each: (collapse, news, politics) and (askreddit, news, politics); and \textit{(iv)} all subreddits combined (see Table~\ref{tab:dataset_stats}).
These particular pairs and triplets were selected based on the number of users participating across the considered subreddits\footnote{See Section 2 in Supplemental Material.}. This approach allowed us to explore whether users' opinions might propagate differently with exposure to multiple subreddit communities.

\smallskip
\subsubsection{Evaluation} \label{subsubsec:eval}
We focused our analysis on users with a sufficient level of engagement, defined as those who contributed to at least 2 events (submissions or comments) with an associated opinion x$_v \in [0,1]$. For users with multiple scored interactions within a simulation interval $s$, we computed their average opinion for that interval. For each user $v$, we then compared their real opinion x$_v$ at time $s$ with their simulated opinion $\hat{x}_v$ at time $s$. We calculated the real opinion x$_v$ as the average of all the user opinions in the interval $s$; the results do not change significantly when considering only the last user opinion in the interval.

To evaluate the results of the diffusion model applied to the three network structures against criteria C1 and C3, we formulated the problem similarly to a (multi-class) classification task. % using standard metrics such as the F1 score. 
This approach required discretizing the continuous opinion vector into distinct categories using some defined thresholds. In our work, we exploited three discretization configurations: \textit{(i)}~binary classification (positive, negative opinions); \textit{(ii)} ternary classification (positive, neutral, negative opinions); and \textit{(iv)} fine-grained classification (positive, toward positive, toward negative, negative opinions). Rather than using fixed thresholds, we implemented an AUC-like approach by exhaustively exploring all possible threshold combinations. For the fine-grained classification, this involved evaluating 84 combinations in the form $[0, \text{thr}_1, \text{thr}_2, \text{thr}_3, 1]$, where $\text{thr}_1 < \text{thr}_2 < \text{thr}_3$ and each threshold varied in the range $[0.1, 0.2, \dots, 0.9]$.
While our analysis employed all three configurations, we primarily discuss results from the fine-grained classification scenario as it provides the most nuanced classification. The reported results represent averages across all threshold combinations. It is worth noting that the observed trends remained consistent across all configurations.

%
%
%
% \vspace{-0.1cm}
\subsection{High-order vs. pairwise opinion diffusion}
In the following, we analyze the performance of the hypergraph-based model, which accounts for group interactions, against the clique-based and graph-based models, which do not incorporate group dynamics. We assess the results of the diffusion model run on top of these models against the defined evaluation criteria to determine the impact of group interaction modeling on predictive accuracy. For clarity, we will directly reference the underlying conversational network model when presenting and discussing the results.
%We assess these models against the defined evaluation criteria to determine the impact of group interaction modeling on predictive accuracy. 

%%%%%%%%%%%%%%%%%%%%%%%%%%%%
% figure 4
%%%%%%%%%%%%%%%%%%%%%%%%%%%%
% acc values sorted by avg size of hyperedges
% C1
\begin{figure}[t!]
    \centering

    \begin{subfigure}{\columnwidth}
        \pgfplotstableread[
    col sep=comma,
]{images/exp_across_nets/data/Criterion1_matched_drifters.csv}\datatable

\begin{tikzpicture}

    % drifted users
    \begin{axis}[
        width=.75\columnwidth, % Scale the plot to \linewidth
        height=6cm,
        xmajorgrids=true,
        grid style={gray!30}, %dashed,
        % xlabel= Simulation steps,
        ytick={1,...,15},
        yticklabels from table={\datatable}{subreddit},
        yticklabel style={
            font=\footnotesize
        },
        ytick style = {
            draw=none
        },
        %xticklabel pos=right,
        % ytick={0,0.2,...,1},
        xlabel={Avg. accuracy}, %\% identified users
        %ylabel=$\ $,
        % ylabel shift={10pt},
        %ylabel style={rotate=-90},
        %yticklabels=\empty,
        %xticklabels=\empty,
        xmin=5, 
        xmax=25, 
        %xmin=0, xmax=1, scaled x ticks=false %, 
        % xtick distance=0.5,
        %ytick distance=0.1,
        x label style={
            font=\small
        },
        xticklabel style={
            font=\small
        },
        xtick style = {
            draw=none
        },
        enlarge y limits = 0.05,
        enlarge x limits = 0.05,
        cycle list={{cerulean},{orange},{darkgreen},{red}},
        mark options={solid, fill, fill opacity=0.65}, %mark repeat={5}
        mark size=2pt,
        legend entries = {Hypergraph, Clique, Graph},
        legend cell align={left},
        legend style={
            at={(0.97, 1.14)},
            legend columns=3,
            column sep=0.5em,
            font=\footnotesize,
        }
    ]
    
        \addplot+[thick, only marks, error bars/.cd, x dir=both, x explicit]
        table[
            y=x,
            x=acc-means,
            col sep=comma,
            x error plus=acc-stds,
            x error minus=acc-stds
        ]{\datatable};
        
        \addplot+[thick,mark=triangle*, only marks, error bars/.cd, x dir=both, x explicit]
        table[
            y=x,
            x=acc-means-cl,
            col sep=comma,
            x error plus=acc-stds-cl,
            x error minus=acc-stds-cl
        ] {\datatable};
        
        \addplot+[thick, mark=diamond*, only marks, error bars/.cd, x dir=both, x explicit] 
        table[
            y=x,
            x=acc-means-gr,
            col sep=comma,
            x error plus=acc-stds-gr,
            x error minus=acc-stds-gr
        ] {\datatable};
    
    \end{axis}

    % \addplot+[
    %         thick, only marks, error bars/.cd, y dir=both, y explicit
    %     ] %, only marks
    %     table[
    %         x=x,
    %         y=avg-mismatched-intervals,
    %         col sep=comma,
    %         y error plus=variance-anticipated-intervals, 
    %         y error minus=variance-anticipated-intervals
    %     ]{\datatable};

\end{tikzpicture}
    \end{subfigure}
    
    \caption{Average accuracy scores for the criterion C1 with standard deviation values.}
    \label{fig:c1_avg_acc}
\end{figure}
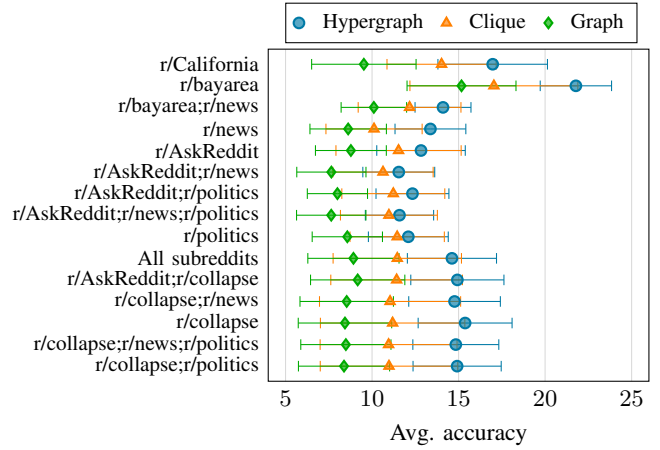

\smallskip
\textit{How well does a model identify a user's first opinion change?}
Fig.~\ref{fig:c1_avg_acc} reports the average accuracy scores for criterion C1, which assesses how well each model detects users' initial opinion shifts. These scores represent the percentage of cases where the model successfully identified both the occurrence and direction (positive or negative) of opinion changes. All datasets in this and subsequent figures are sorted according to the average hyperedge size per simulation interval.

The hypergraph model demonstrates better performance, consistently achieving higher accuracy scores compared to both clique-based and graph representations across almost all scenarios. Statistical analysis reveals statistically significant differences (p=0.05) between all results, except between the scores of the hypergraph and clique models in r/politics and in the combined dataset of r/AskReddit, r/news, and r/politics.

A notable pattern emerges in datasets containing r/politics and r/AskReddit, where we observe a marked decrease in performance. This anomaly likely stems from the nature of these communities. In particular, r/AskReddit is described as a ``place to ask and answer thought-provoking questions", where comments potentially exhibit more rhetorical structures, such as irony, sarcasm, or subtle argumentative structures. We found no significant correlation between this performance drop and any particular structural properties of the conversational (hyper-)networks.

\begin{figure}[b!]
    \centering

    \begin{subfigure}{\columnwidth}
        \pgfplotstableread[
    col sep=comma,
]{images/exp_across_nets/data/Criterion2_direction.csv}\datatable

\pgfplotsset{
    scatter/classes={%
        AskReddit={mark=square*},
        bayarea={mark=square},%
        California={mark=triangle*},%
        collapse={mark=o},%
        news={mark=diamond*},%
        politics={mark=pentagon*},%
        AskRedditCollapse={mark=asterisk},
        AskRedditNews={mark=10-pointed star},
        AskRedditPolitics={mark=Mercedes star},
        bayareaNews={mark=star},
        collapseNews={mark=pentagon},
        collapsePolitics={mark=halfcircle*},
        AskRedditNewsPolitics={mark=otimes},
        collapseNewsPolitics={mark=oplus},
        all={mark=|}
        % $DynamicGreedy_{[H]_2}-noOpt$={mark=halfsquare*,cyan,mark options={cyan,}},%scale=2
        % SubTSS={mark=pentagon,draw=darkgreen},%
        % SubTSS-noOpt={mark=star,draw=darkyellow}%
    },
}

\begin{tikzpicture}
    \begin{axis}[
        width=.75\columnwidth, % Scale the plot to \linewidth
        height=6cm,
        grid=major,
        grid style={dashed,gray!30},
        xlabel={\small Avg. postponed intervals},
        %xtick={1,...,15},
        %xticklabels from table={\datatable}{subreddit},
        xticklabel style={
            font=\small
        },
        %xticklabel pos=right,
        % ytick={0,0.2,...,1},
        ylabel={Avg. anticipated intervals},
        %ylabel=$\ $,
        % ylabel shift={10pt},
        %ylabel style={rotate=-90},
        %yticklabels=\empty,
        %xticklabels=\empty,
        % xmin=20,
        % ymin=20, 
        % ymax=60, 
        %xmin=0, xmax=1, scaled x ticks=false %, 
        % xtick distance=0.5,
        %ytick distance=0.1,
        y label style={
            font=\small
        },
        yticklabel style={
            font=\small
        },
        cycle list={{gray},{cerulean},{orange},{darkgreen},{red}},
        mark options={solid, fill, fill opacity=0.65}, %mark repeat={5}
        mark size=2pt,
        legend entries = {
            r/AskReddit,
            r/bayarea,
            r/California,
            r/collapse,
            r/news,
            r/politics,
            r/AskReddit;r/collapse,
            r/AskReddit;r/news,
            r/AskReddit;r/politics,
            r/bayarea;r/news,
            r/collapse;r/news,
            r/collapse;r/politics,
            r/AskReddit;r/news;r/politics,
            r/collapser/news;r/politics,
            All subreddits
        },
        legend cell align={left},
        legend style={
            at={(1.6, 1.0)},
            legend columns=1,
            column sep=0.5em,
            font=\tiny,
        }
    ]

        % adding this just to have a gray color on the legend
        \addplot+[only marks, scatter, scatter src=explicit symbolic]
        table[
            x=avg-postponed-intervals,
            y=avg-anticipated-intervals,
            col sep=comma,
            meta=class
        ]{\datatable};
    
        \addplot+[only marks, scatter, scatter src=explicit symbolic]
        table[
            x=avg-postponed-intervals,
            y=avg-anticipated-intervals,
            col sep=comma,
            meta=class
        ]{\datatable};

        \addplot+[only marks, scatter, scatter src=explicit symbolic]
        table[
            x=avg-postponed-intervals-cl,
            y=avg-anticipated-intervals-cl,
            col sep=comma,
            meta=class
        ]{\datatable};

        \addplot+[only marks, scatter, scatter src=explicit symbolic]
        table[
            x=avg-postponed-intervals-gr,
            y=avg-anticipated-intervals-gr,
            col sep=comma,
            meta=class
        ]{\datatable};

        % adding a diagonal line
        \draw[lightgray, thick] (0,0)  -- (80,80);

    \end{axis}
\end{tikzpicture}
    \end{subfigure}
    \caption{Avg. anticipated vs. postponed simulation intervals in identifying the first opinion drift. {Hypergraph} results are shown in {\color{cerulean}\textbf{blue}}, {clique} results in {\color{orange}\textbf{orange}}, and {graph} results in {\color{darkgreen}\textbf{green}}.}
    \label{fig:c2_scatter}
\end{figure}
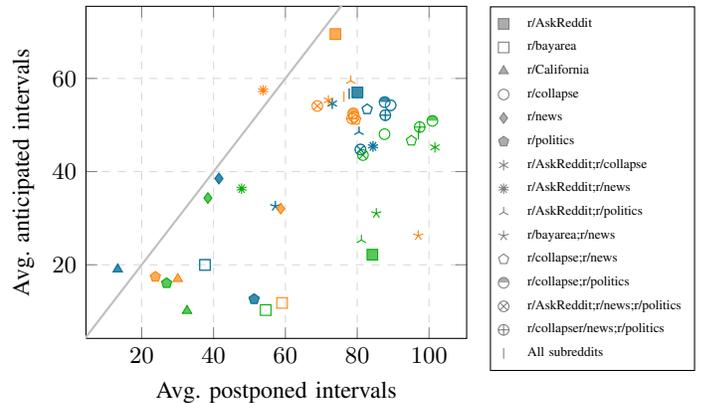

\begin{figure}[t!]
    \centering

    \begin{subfigure}{\columnwidth}
        \pgfplotstableread[
    col sep=comma,
]{images/exp_across_nets/data/Criterion2_direction.csv}\datatable

\begin{tikzpicture}
    \begin{axis}[
        width=.75\columnwidth, % Scale the plot to \linewidth
        height=6cm,
        xmajorgrids=true,
        grid style={gray!30}, %dashed,
        % xlabel= Simulation steps,
        ytick={1,...,15},
        yticklabels from table={\datatable}{label},
        yticklabel style={
            font=\footnotesize
        },
        ytick style = {
            draw=none
        },
        %xticklabel pos=right,
        % ytick={0,0.2,...,1},
        xlabel={Average RMSE},
        %ylabel=$\ $,
        % ylabel shift={10pt},
        %ylabel style={rotate=-90},
        %yticklabels=\empty,
        %xticklabels=\empty,
        % ymin=0, 
        % ymax=0.5, 
        %xmin=0, xmax=1, scaled x ticks=false %, 
        % xtick distance=0.5,
        %ytick distance=0.1,
        x label style={
            font=\small
        },
        xticklabel style={
            font=\small
        },
        xtick style = {
            draw=none
        },
        enlarge y limits = 0.05,
        enlarge x limits = 0.05,
        cycle list={{cerulean},{orange},{darkgreen},{red}},
        mark options={solid, fill, fill opacity=0.65}, %mark repeat={5}
        mark size=2pt,
        legend entries = {Hypergraph, Clique, Graph},
        legend cell align={left},
        legend style={
            at={(0.97, 1.14)},
            legend columns=3,
            column sep=0.5em,
            font=\footnotesize,
        }
    ]
    
        \addplot+[
            thick, only marks, error bars/.cd, x dir=both, x explicit
        ] %, only marks
        table[
            y=x,
            x=avg-mismatched-intervals,
            col sep=comma,
            x error plus=std-anticipated-intervals, 
            x error minus=std-anticipated-intervals
        ]{\datatable};
        
        \addplot+[
            thick,mark=triangle*, only marks, error bars/.cd, x dir=both, x explicit
        ]
        table[
            y=x,
            x=avg-mismatched-intervals-cl,
            col sep=comma,
            x error plus=std-anticipated-intervals-cl, 
            x error minus=std-anticipated-intervals-cl
        ] {\datatable};
        
        \addplot+
        [
            thick, mark=diamond*, only marks, error bars/.cd, x dir=both, x explicit
        ] 
        table[
            y=x,
            x=avg-mismatched-intervals-gr,
            col sep=comma,
            x error plus=std-anticipated-intervals-gr, 
            x error minus=std-anticipated-intervals-gr
        ] {\datatable};
    
    \end{axis}
\end{tikzpicture}
    \end{subfigure}
    
    \caption{Average error in identifying the first opinion drift with standard deviation values.}
    \label{fig:c2_error}
\end{figure}
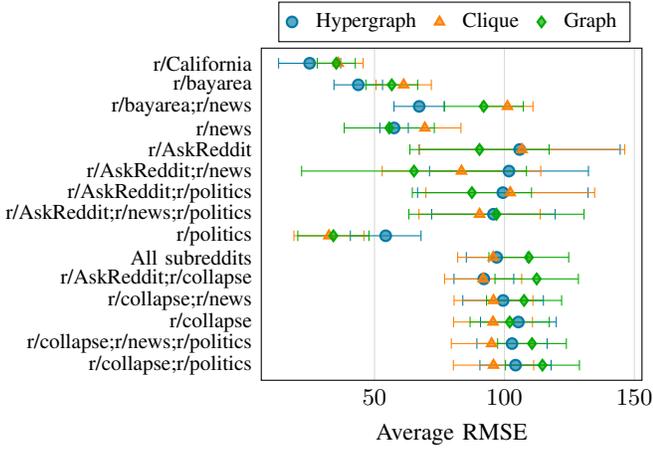

\medskip
\textit{Does a model tend to anticipate or postpone a user's opinion drift?}
Our analysis reveals no consistent pattern across datasets regarding whether the models anticipate or postpone users' opinion drifts during the simulation. Contrary to initial expectations that the hypergraph model might consistently anticipate opinion changes due to the compounded influence of groups on individuals, the empirical results demonstrate a more complex reality. As illustrated in Fig.~\ref{fig:c2_scatter}, all models tend to postpone rather than anticipate users' opinion drifts, though with some variations. Specifically, the clique-based model (orange points) shows a greater tendency to anticipate drifts more frequently across the datasets, while the graph-based model (green points) consistently leans towards postponing opinion changes.
When focusing on the average error in identifying a user’s first opinion drift (see Fig.~\ref{fig:c2_error}), we still observe an absence of a consistent pattern across datasets. 

% C2

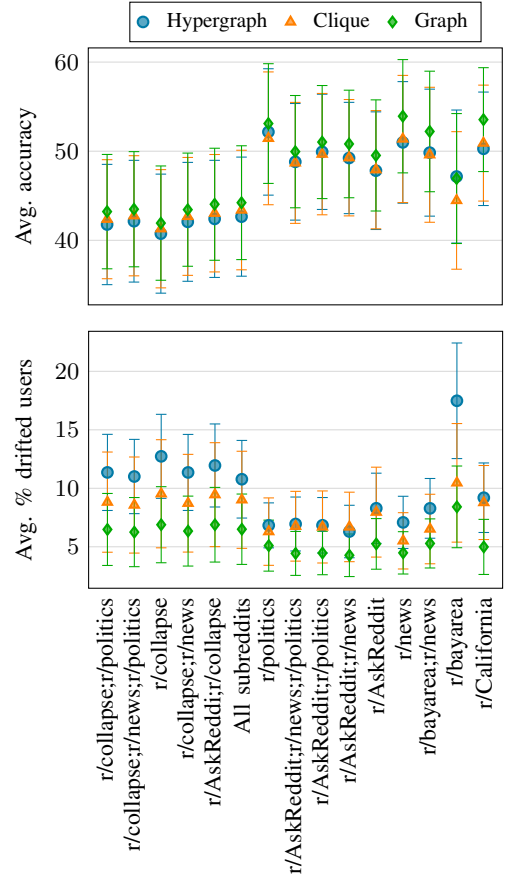
\begin{figure}
    \centering
    \begin{subfigure}{.8\columnwidth}
        \pgfplotstableread[
    col sep=comma,
]{images/exp_across_nets/data/Criterion3.csv}\datatable

\begin{tikzpicture}

    % drifted users
    \begin{axis}[
        width=\columnwidth, % Scale the plot to \linewidth
        height=5cm,
        ymajorgrids=true,
        grid style={gray!30}, %dashed,
        % xlabel= Simulation steps,
        xtick={1,...,15},
        % xticklabels from table={\datatable}{subreddit},
        xticklabel style={
            font=\small,
            opacity=0,
        },
        xtick style = {
            draw=none
        },
        %xticklabel pos=right,
        % ytick={0,0.2,...,1},
        ylabel={Avg. accuracy}, %\% matched users
        %ylabel=$\ $,
        % ylabel shift={10pt},
        %ylabel style={rotate=-90},
        %yticklabels=\empty,
        %xticklabels=\empty,
        % ymin=0, 
        % ymax=0.5, 
        %xmin=0, xmax=1, scaled x ticks=false %, 
        % xtick distance=0.5,
        %ytick distance=0.1,
        y label style={
            font=\small
        },
        yticklabel style={
            font=\small
        },
        ytick style = {
            draw=none
        },
        enlarge y limits = 0.05,
        enlarge x limits = 0.05,
        cycle list={{cerulean},{orange},{darkgreen},{red}},
        mark options={solid, fill, fill opacity=0.65}, %mark repeat={5}
        mark size=2pt,
        legend entries = {Hypergraph, Clique, Graph},
        legend cell align={left},
        legend style={
            at={(0.95, 1.18)},
            legend columns=3,
            column sep=0.5em,
            font=\footnotesize,
        }
    ]
    
        \addplot+[thick, only marks, error bars/.cd, y dir=both, y explicit]
        table[
            x=x,
            y=avg-matched-users,
            col sep=comma,
            y error plus=variance-matched-users,
            y error minus=variance-matched-users
        ]{\datatable};
        
        \addplot+[thick,mark=triangle*, only marks, error bars/.cd, y dir=both, y explicit]
        table[
            x=x,
            y=avg-matched-users-cl,
            col sep=comma,
            y error plus=variance-matched-users-cl,
            y error minus=variance-matched-users-cl
        ] {\datatable};
        
        \addplot+[thick, mark=diamond*, only marks, error bars/.cd, y dir=both, y explicit] 
        table[
            x=x,
            y=avg-matched-users-gr,
            col sep=comma,
            y error plus=variance-matched-users-gr,
            y error minus=variance-matched-users-gr
        ] {\datatable};
    
    \end{axis}

\end{tikzpicture}
    \end{subfigure}

    \vspace*{-0.1cm}
    % among the users correctly identified, 
    % How many have changed their opinion from the beginning?
    \begin{subfigure}{.8\columnwidth}
        \pgfplotstableread[
    col sep=comma,
]{images/exp_across_nets/data/Criterion3.csv}\datatable

\begin{tikzpicture}

    % drifted users
    \begin{axis}[
        width=\columnwidth, % Scale the plot to \linewidth
        height=5cm,
        ymajorgrids=true,
        grid style={gray!30}, %dashed,
        % xlabel= Simulation steps,
        xtick={1,...,15},
        xticklabels from table={\datatable}{subreddit},
        xticklabel style={
            rotate=90,
            font=\small
        },
        xtick style = {
            draw=none
        },
        %xticklabel pos=right,
        % ytick={0,0.2,...,1},
        ylabel={Avg. \% drifted users},
        %ylabel=$\ $,
        % ylabel shift={10pt},
        %ylabel style={rotate=-90},
        %yticklabels=\empty,
        %xticklabels=\empty,
        % ymin=0, 
        % ymax=0.5, 
        %xmin=0, xmax=1, scaled x ticks=false %, 
        % xtick distance=0.5,
        %ytick distance=0.1,
        y label style={
            font=\small
        },
        yticklabel style={
            font=\small
        },
        ytick style = {
            draw=none
        },
        enlarge y limits = 0.05,
        enlarge x limits = 0.05,
        cycle list={{cerulean},{orange},{darkgreen},{red}},
        mark options={solid, fill, fill opacity=0.65}, %mark repeat={5}
        mark size=2pt,
        % legend entries = {Hypergraph, Clique, Graph},
        % legend cell align={left},
        % legend style={
        %     at={(0.8, 0.2)},
        %     % anchor=north, 
        %     legend columns=3,
        %     column sep=0.5em,
        %     font=\small,
        %     % draw=none
        % }
    ]
    
        \addplot+[thick, only marks, error bars/.cd, y dir=both, y explicit]
        table[
            x=x,
            y=avg-drifted-users,
            col sep=comma,
            y error plus=variance-drifted-users,
            y error minus=variance-drifted-users
        ]{\datatable};
        
        \addplot+[thick,mark=triangle*, only marks, error bars/.cd, y dir=both, y explicit]
        table[
            x=x,
            y=avg-drifted-users-cl,
            col sep=comma,
            y error plus=variance-drifted-users-cl,
            y error minus=variance-drifted-users-cl
        ] {\datatable};
        
        \addplot+[thick, mark=diamond*, only marks, error bars/.cd, y dir=both, y explicit] 
        table[
            x=x,
            y=avg-drifted-users-gr,
            col sep=comma,
            y error plus=variance-drifted-users-gr,
            y error minus=variance-drifted-users-gr
        ] {\datatable};
    
    \end{axis}

    % \addplot+[
    %         thick, only marks, error bars/.cd, y dir=both, y explicit
    %     ] %, only marks
    %     table[
    %         x=x,
    %         y=avg-mismatched-intervals,
    %         col sep=comma,
    %         y error plus=variance-anticipated-intervals, 
    %         y error minus=variance-anticipated-intervals
    %     ]{\datatable};

\end{tikzpicture}
    \end{subfigure}
    
    \caption{Top plot: Average accuracy scores for criterion C3, including standard deviation values. Bottom plot: Average percentage of users who changed their opinion from the beginning to the end, considering only those correctly identified.}
    \label{fig:c3}
\end{figure}

\smallskip
\textit{How well does each model predict a user's final stance?}
Fig.~\ref{fig:c3} (top plot) reports the average accuracy scores for criterion C3, which evaluates the models' ability to detect users' final opinions. Our analysis indicates that, in most cases, there are no statistically significant differences between models regarding their accuracy in predicting final user stances. The only notable exceptions appear in the last four datasets, where the graph model demonstrates a modest advantage, correctly identifying up to 3\% more users.

However, a key distinction emerges when examining the models' performance by focusing on users who modified their initial opinions among those correctly identified (see Fig.~\ref{fig:c3}, bottom plot). In this case, the hypergraph model consistently demonstrates superior accuracy in detecting final opinions. This performance advantage holds across most datasets, with the exception of those spanning from r/politics through r/AskReddit and r/California, where the clique-based representation achieves comparable performance to the hypergraph~model.

% C3

\subsection{Investigating the effect of time}
\begin{figure}
    \centering

    \begin{subfigure}{.5\textwidth}
        \begin{tikzpicture}
    \begin{axis}[
        %width=.9\columnwidth, % Scale the plot to \linewidth
        height=5.5cm,
        grid=major,
        grid style={dashed,gray!30},
        xlabel = Median hyperedge size,
        ylabel = {CCDF}, %Sample with value $>$ Degree
        ylabel near ticks,
        xlabel near ticks, 
        ylabel style = {
            font=\small
        },
        xlabel style = {
            font=\small
        },
        yticklabel style = {
            font=\footnotesize,
            xshift=0.5ex
        },
        xticklabel style = {
            font=\footnotesize,
            yshift=0.5ex
        },
        title style = {
            font=\small,
            yshift=-0.5ex
        },
        cycle list={{cerulean},{orange},{darkgreen},{red}},
        mark options={solid, fill, fill opacity=0.65}, 
        mark size=2pt,
        legend entries = {
            $\Delta$ = 6h,
            $\Delta$ = 12h,
            $\Delta$ = 24h,
            $\Delta$ = 1 week,
        },
        % legend pos=south east,
        legend cell align={left},
        legend style={
            at={(1.55, 1.)},
            legend columns=1,
            % column sep=0.5em,
            font=\footnotesize,
        },
        % xmode=log,
        ymode=log
    ]
    
        \addplot+[thick, mark=*]
        table[x=size,y=cs,col sep=comma] {./images/exp_across_time/data/ccdf/hour_ccdf.csv};

        \addplot+[thick, mark=triangle*]
        table[x=size,y=cs,col sep=comma] {./images/exp_across_time/data/ccdf/halfday_ccdf.csv};

        \addplot+[thick, mark=diamond*]
        table[x=size,y=cs,col sep=comma] {./images/exp_across_time/data/ccdf/day_ccdf.csv};

        \addplot+[thick, mark=star]
        table[x=size,y=cs,col sep=comma] {./images/exp_across_time/data/ccdf/week_ccdf.csv};
        
        % \addplot
        % table[x=DEG,y=CS,col sep=comma, mark=square,
        % samples=20] {./images/networks/data/ccdf_before_backbone.txt};
        
    \end{axis}
\end{tikzpicture}
    \end{subfigure}

    %\vspace{-0.5cm}
    
    \caption{Comparison of the Complementary Cumulative Distribution Function (CCDF) for the hyperedge size distributions across the four temporal hyper-conversational networks, obtained considering four distinct time windows.}
    \label{fig:ccdf_indegree}
\end{figure}
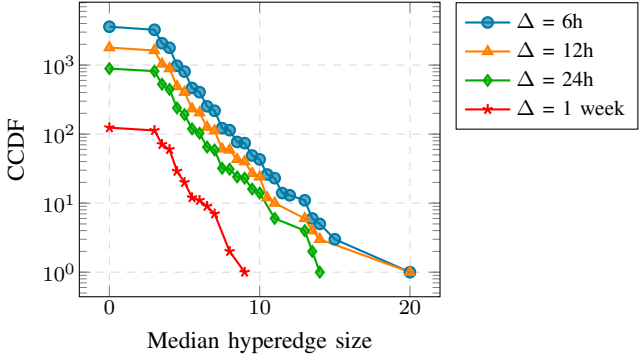

% here we detail that
% - we ranged the size of the discretization window
% - we used a single aggregated dataset
The dynamics of online social influence and opinion change are inherently linked to exogenous events that may stimulate engaged discussions on social media platforms~\cite{Eminente_Chaos_2022,Gong_JoC_2023}. When studying opinion propagation in these environments, it is crucial to accurately define the timeframe during which a comment can effectively influence its readers. 
Our second experiment focuses on this aspect and examines how different configurations of the time window parameter ($\Delta$) affect opinion spreading within the network as it controls which comments are included in the conversational (hyper-)network structure. We evaluated four distinct time windows: 6 hours, 12 hours, 24 hours, and one week (168 hours). Fig.~\ref{fig:ccdf_indegree} shows the distribution of the median hyperedge size per temporal interval for the four values of $\Delta$. Interestingly, despite the intuition that larger time windows would capture more complete conversations and reveal larger hyperedges by allowing more users to contribute to the same subthread, our analysis shows the opposite effect—median hyperedge size actually decreases with larger windows. This pattern reflects the bursty nature of Reddit conversations, where communication activity tends to concentrate in short, intense periods rather than distributing evenly over time~\cite{Desiderio_pnasnexus_2025}.
In this analysis, we used the same experimental framework outlined in Section~\ref{subsec:exp_setting}, and ran this experiment considering all subreddits as a single dataset. 

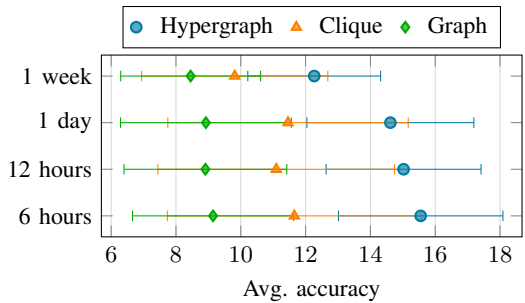
\begin{figure}[b!]
    \centering

    \begin{subfigure}{.9\columnwidth}
        \pgfplotstableread[
    col sep=comma,
]{images/exp_across_time/data/Criterion1_matched_drifters_time.csv}\datatable

\begin{tikzpicture}

    % drifted users
    \begin{axis}[
        width=.9\columnwidth, % Scale the plot to \linewidth
        height=4cm,
        xmajorgrids=true,
        grid style={gray!30}, %dashed,
        % xlabel= Simulation steps,
        ytick={1,...,4},
        yticklabels from table={\datatable}{label},
        yticklabel style={
            font=\small
        },
        %xticklabel pos=right,
        % ytick={0,0.2,...,1},
        xlabel={Avg. accuracy}, %\% identified users
        %ylabel=$\ $,
        % ylabel shift={10pt},
        %ylabel style={rotate=-90},
        %yticklabels=\empty,
        %xticklabels=\empty,
        % xmin=5, 
        % xmax=25, 
        %xmin=0, xmax=1, scaled x ticks=false %, 
        % xtick distance=0.5,
        %ytick distance=0.1,
        x label style={
            font=\small
        },
        xticklabel style={
            font=\small
        },
        enlarge y limits = 0.15,
        enlarge x limits = 0.05,
        cycle list={{cerulean},{orange},{darkgreen},{red}},
        mark options={solid, fill, fill opacity=0.65}, %mark repeat={5}
        mark size=2pt,
        legend entries = {Hypergraph, Clique, Graph},
        legend cell align={left},
        legend style={
            at={(0.95, 1.27)},
            legend columns=3,
            column sep=0.5em,
            font=\small,
        }
    ]
    
        \addplot+[thick, only marks, error bars/.cd, x dir=both, x explicit]
        table[
            y=x,
            x=acc-means,
            col sep=comma,
            x error plus=acc-stds,
            x error minus=acc-stds
        ]{\datatable};
        
        \addplot+[thick,mark=triangle*, only marks, error bars/.cd, x dir=both, x explicit]
        table[
            y=x,
            x=acc-means-cl,
            col sep=comma,
            x error plus=acc-stds-cl,
            x error minus=acc-stds-cl
        ] {\datatable};
        
        \addplot+[thick, mark=diamond*, only marks, error bars/.cd, x dir=both, x explicit] 
        table[
            y=x,
            x=acc-means-gr,
            col sep=comma,
            x error plus=acc-stds-gr,
            x error minus=acc-stds-gr
        ] {\datatable};
    
    \end{axis}

    % \addplot+[
    %         thick, only marks, error bars/.cd, y dir=both, y explicit
    %     ] %, only marks
    %     table[
    %         x=x,
    %         y=avg-mismatched-intervals,
    %         col sep=comma,
    %         y error plus=variance-anticipated-intervals, 
    %         y error minus=variance-anticipated-intervals
    %     ]{\datatable};

\end{tikzpicture}
    \end{subfigure}
    
    \caption{Average accuracy scores for the criterion C1 with standard deviation values.}
    \label{fig:c1_avg_acc_time}
\end{figure}

\smallskip
\textit{What is the impact of the time window parameter $\Delta$ on detecting initial opinion change?}
Fig.~\ref{fig:c1_avg_acc_time} presents the average accuracy scores for criterion C1, revealing two significant findings regarding the impact of time window configuration on detecting users' first opinion shifts. First, the analysis demonstrates a consistent pattern across all scenarios: the hypergraph model exhibits statistically significant better performance compared to both clique-based and graph models (p=0.05). The second key finding relates to temporal granularity: the hypergraph and clique-based models show diminished performance when analyzing user opinions across broader time windows (one week in this case). 

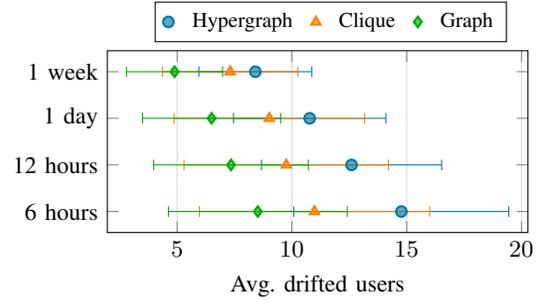
\begin{figure}[t!]
    \centering

    \begin{subfigure}{.9\columnwidth}
        \pgfplotstableread[
    col sep=comma,
]{images/exp_across_time/data/Criterion3.csv}\datatable

\begin{tikzpicture}

    % drifted users
    \begin{axis}[
        width=.9\columnwidth, % Scale the plot to \linewidth
        height=4cm,
        xmajorgrids=true,
        grid style={gray!30}, %dashed,
        % xlabel= Simulation steps,
        ytick={1,...,4},
        yticklabels from table={\datatable}{label},
        yticklabel style={
            font=\small
        },
        %xticklabel pos=right,
        % ytick={0,0.2,...,1},
        xlabel={Avg. drifted users}, %\% identified users
        %ylabel=$\ $,
        % ylabel shift={10pt},
        %ylabel style={rotate=-90},
        %yticklabels=\empty,
        %xticklabels=\empty,
        % xmin=5, 
        % xmax=25, 
        %xmin=0, xmax=1, scaled x ticks=false %, 
        % xtick distance=0.5,
        %ytick distance=0.1,
        x label style={
            font=\small
        },
        xticklabel style={
            font=\small
        },
        ytick style = {
        },
        enlarge y limits = 0.15,
        enlarge x limits = 0.05,
        cycle list={{cerulean},{orange},{darkgreen},{red}},
        mark options={solid, fill, fill opacity=0.65}, %mark repeat={5}
        mark size=2pt,
        legend entries = {Hypergraph, Clique, Graph},
        legend cell align={left},
        legend style={
            at={(0.95, 1.27)},
            legend columns=3,
            column sep=0.5em,
            font=\footnotesize,
        }
    ]
    
        \addplot+[thick, only marks, error bars/.cd, x dir=both, x explicit]
        table[
            y=x,
            x=avg-drifted-users,
            col sep=comma,
            x error plus=std-drifted-users,
            x error minus=std-drifted-users
        ]{\datatable};
        
        \addplot+[thick,mark=triangle*, only marks, error bars/.cd, x dir=both, x explicit]
        table[
            y=x,
            x=avg-drifted-users-cl,
            col sep=comma,
            x error plus=std-drifted-users-cl,
            x error minus=std-drifted-users-cl
        ] {\datatable};
        
        \addplot+[thick, mark=diamond*, only marks, error bars/.cd, x dir=both, x explicit] 
        table[
            y=x,
            x=avg-drifted-users-gr,
            col sep=comma,
            x error plus=std-drifted-users-gr,
            x error minus=std-drifted-users-gr
        ] {\datatable};
    
    \end{axis}

    % \addplot+[
    %         thick, only marks, error bars/.cd, y dir=both, y explicit
    %     ] %, only marks
    %     table[
    %         x=x,
    %         y=avg-mismatched-intervals,
    %         col sep=comma,
    %         y error plus=variance-anticipated-intervals, 
    %         y error minus=variance-anticipated-intervals
    %     ]{\datatable};

\end{tikzpicture}
    \end{subfigure}
    
    \caption{
    Average percentage of users who changed their opinion from the beginning to the end, considering only those correctly identified.}
    \label{fig:c3_avg_acc_time}
\end{figure}

\smallskip
\textit{What is the impact of the time window parameter $\Delta$ on detecting final opinions?}
Fig.~\ref{fig:c3_avg_acc_time} presents the average accuracy scores for criterion C3, focusing on the percentage of correctly identified users who changed their initial opinions. The results follow a pattern consistent with the findings of criterion C1. Across all temporal configurations, the hypergraph model outperforms both the clique-based and graph models in correctly detecting final opinions. Nevertheless, similar to our previous observations, the ability to accurately predict final opinions decreases significantly when considering wider time windows.

%%%%%%%%%%%%%%%%%%%%%%%%%%%%%%%%%%%%%%%%%%%%%%%%%%%%%%%%%%%%%
% DISCUSSIONS
%%%%%%%%%%%%%%%%%%%%%%%%%%%%%%%%%%%%%%%%%%%%%%%%%%%%%%%%%%%%%
\section{Discussion}\label{sec:discussion}

\textit{
    Modeling conversational networks using (temporal) hypergraphs consistently enables more accurate identification of users' initial opinion shifts, regardless of the size of the time window used to model the influence of comments on the readers. Our analysis found no clear patterns indicating whether the simulated opinion shifts tended to occur before or after the observed shifts.
}

\smallskip

Analysis of accuracy scores for criterion C1 across all datasets (see Fig.~\ref{fig:c1_avg_acc}) and time window parameter $\Delta$ (see Fig.~\ref{fig:c1_avg_acc_time}) demonstrates that the hypergraph model consistently surpasses both the graph model and, in most cases, the clique-based model in performance. This indicates that the hypergraph model is more effective at predicting the first opinion drift for a larger percentage of users, regardless of the granularity of the observed interactions.
These findings highlight the advantages of modeling conversational networks with high-order structures, as the hypergraph model captures subtle group dynamics within conversational threads that other models may overlook. This advantage likely arises from the inherent nature of the hypergraph model, which accounts for the cumulative effects of group interactions on individual opinions. In contrast, the graph model only considers direct user-to-user interactions, while the clique model, though accounting for group interactions, limits its scope to pairwise relationships within groups rather than their cumulative effects, hence potentially failing to fully capture the influence of groups and information exposure on opinion dynamics.

Our analysis reveals that performance generally decreases with broader time windows, indicating that extended sampling periods may complicate the tracking of opinion evolution. This degradation may occur because averaging opinions over longer periods can dilute the representation of a user's true stance, while relying on only the most recent opinion risks missing significant opinion shifts. These findings emphasize the critical importance of precisely defining the temporal relevance of comments—the period during which they maintain their potential to influence readers—as broader time windows can lead to less accurate representations of user opinions and subsequently less reliable predictions of opinion shifts.

Contrary to initial expectations, we found no consistent pattern in the timing of the simulated opinion changes across models. One might intuitively assume that incorporating group interactions with the hypergraph model would lead to an earlier prediction of opinion changes due to the compounded influence of group dynamics, while the graph model, focused on pairwise interactions, would be expected to postpone opinion shifts. While this expectation holds somehow true for the latter model, the former exhibits different behavior than expected (see Fig.~\ref{fig:c2_scatter}). Specifically, these results suggest that while modeling group dynamics may lead to more opinion changes, it does not necessarily accelerate the timing of these changes.

\bigskip
\textit{Modeling conversational networks with hypergraphs allows for a consistently more accurate prediction of users' final opinions when considering those who changed their initial stance, regardless of the size of the time window used to model the influence of comments on the readers.}

\begin{table}[b!]
    \setlength\extrarowheight{1pt}
    \centering 

    \caption{Zoom in the results of criterion C3 for the combined dataset using a 24-hour time window.}
    \label{tab:c3_results_breakdown}

    \begin{tabular}{lrrr}
        \toprule 

        \textbf{} & 
        \multicolumn{1}{c}{\textbf{Hypegraph}} & 
        \multicolumn{1}{c}{\textbf{Clique}} & 
        \multicolumn{1}{c}{\textbf{Graph}} 
        \\

        \midrule

        \textbf{P$_1$} & 20.59 + 4.85 & 17.17 + 5.86 & 12.91 + 4.24 \\

        \rowcolor{gray!10}
        \textbf{P$_2$} & 79.41 + 4.85 & 82.83 + 5.86 & 87.1 + 4.24 \\

        \textbf{P$_3$} & 8.05 + 1.41  & 6.75 + 2.29 & 4.98 + 1.70 \\

        \rowcolor{gray!10}
        \textbf{P$_4$} & 10.77 + 3.32 & 9.02 + 4.15 & 6.50 + 3.00 \\

        \textbf{P$_5$} & 82.75 + 7.11 & 85.88 + 7.49 & 89.92 + 5.41 \\

        \rowcolor{gray!10}
        \textbf{P$_6$} & 42.66 + 6.68 & 43.40 + 6.71 & 44.22 + 6.39 \\

        \bottomrule
    \end{tabular}
\end{table}

\smallskip

Our evaluation of how effectively different conversational network models predict users' final opinions (criterion C3) reveals two distinct patterns. Initial analysis shows that all models perform similarly when considering the overall user population (Fig.~\ref{fig:c3}, top plot).
%(Figs.~\ref{fig:c3} and~\ref{fig:c3_avg_acc_time}, top plots). 
However, a more focused examination of users who modified their initial opinions yields markedly different results (Fig.~\ref{fig:c3}, bottom plot, and Fig.~\ref{fig:c3_avg_acc_time}). 
%(Figs.~\ref{fig:c3} and~\ref{fig:c3_avg_acc_time}, bottom plots).
In these cases, the hypergraph model demonstrates superior performance compared to both clique-based and graph models in predicting final opinions, maintaining this advantage regardless of interaction granularity. This finding aligns with our earlier observations for criterion C1.

This apparent discrepancy in performance can be explained by examining each model's ability to detect opinion changes. Table~\ref{tab:c3_results_breakdown} details the percentage of:
\textit{(P$_1$)} users who modified their opinions from the beginning of the simulation to the end, 
\textit{(P$_2$)} users who maintained their initial stance,
\textit{(P$_3$)} correctly identified users who changed their opinions within the subset of all users who modified their stance,
\textit{(P$_4$)} correctly identified users who changed their opinions among all correctly classified users (values shown in Fig.~\ref{fig:c3}, bottom plot, and Fig.~\ref{fig:c3_avg_acc_time}), 
\textit{(P$_5$)} correctly identified users who did not modify their stance, and 
\textit{(P$_6$)} the overall match rate between predicted and actual outcomes (values shown in Fig.~\ref{fig:c3}, top plot).
%This data refer to the combined dataset using a 24-hour time window $\Delta$.
%
The table highlights the distinct patterns that the three models simulate in the opinion diffusion process. Specifically, the hypergraph model simulates the highest percentage of users who modified their initial positions (P$_1$), while the graph model the lowest. Conversely, for users maintaining their initial opinions (P$_2$), we observe an inverse relationship. However, all models' predictions deviate significantly from the ground truth (P$_3$--P$_6$), where 54.03\% ± 6.02\% of users changed their opinions and 45.97\% ± 6.02\% maintained their initial positions.
In general, all models demonstrate a higher accuracy rate in predicting users who maintained their initial opinions (P$_5$) compared to those who changed their stance (P$_3$). These findings suggest two key insights. First, predicting opinion stability appears to be inherently easier than predicting opinion changes across all models. Second, group dynamics seem to be valuable indicators of potential opinion shifts. Overall, the hypergraph-based model shows higher sensitivity in detecting opinion fluctuations without significantly compromising its ability to identify opinion stability. 

\medskip
\textit{Answering RQ$_1$ and RQ$_2$.}
% RQ1: How can we capture high-order interactions happening in conversational data?
% RQ2: Does introducing high-order social dynamics into a diffusion model lead to a more realistic process? If so, to what extent does introducing high-order social dynamics enhance our understanding of complex social interactions?
The presence of high-order interactions in social systems has been largely studied in sociology and psychology, showing that group dynamics play a crucial role in shaping individual opinions and behaviors~\cite{Moscovici_EJSP_1976,Turner_1987}. As a consequence, understanding group membership is fundamental for revealing personal behaviors~\cite{Forsyth2019}, as these are strongly influenced by the crowd~\cite{Edelson_Science_2011}. While groups traditionally formed offline, the rise of online social networks has shifted many group decision processes to digital spaces~\cite{Lucchini_RoyalSoc_2022,lenti2024causal}. This transition has amplified the importance of social influence due to the exponential increase in potential user connections. In this context, analyzing online conversations has become essential for understanding offline behavioral patterns.

The research on opinion diffusion models accounting for group dynamics is still in early stages (see Section~\ref{sec:rel_work} - Related Work). However, the lack of ground-truth stance data has hindered these models' ability to replicate real-world opinion dynamics accurately. Our work contributes to this line of research by providing empirical evidence for both the presence and significance of high-order interactions in online discussions.
Our findings address two key research questions. First, regarding RQ$_1$, the higher performance of our temporal hypergraph model, which abstracts conversational threads, suggests its ability to capture subtle group dynamics that form during topic discussions. This outcome indicates that modeling specific conversational dynamics within local segments of conversations can lead to a better understanding of opinion evolution. Second, addressing RQ$_2$, our results suggest that incorporating high-order social dynamics produces more realistic diffusion modeling. Specifically, the hypergraph model consistently outperforms graph-based approaches in predicting both initial opinion changes (3.53\%-7.45\% improvement, Fig.~\ref{fig:c1_avg_acc}) and final opinions of users who modified their views (1.73\%-9.07\% improvement, Fig.~\ref{fig:c3} bottom) across all datasets. This advantage persists across various time window configurations, with the hypergraph model showing greater sensitivity in detecting first opinion changes (3.81\%-6.41\% improvement, Fig.~\ref{fig:c1_avg_acc_time}) and more accurately predicting final opinions (3.52\%-6.25\% improvement, Fig.~\ref{fig:c3_avg_acc_time}). Importantly, we found that the granularity of the observation of these interactions significantly impacts prediction accuracy.

These results represent an important step toward bridging the gap between theoretical models of opinion dynamics and real-world validation. We empirically showed that high-order interactions significantly influence opinion dynamics in online discussions, with evidence of group dynamics affecting opinion formation across different subreddits. However, we must acknowledge a critical aspect: all models' predictions deviate substantially from our synthetic ground-truth data, indicating that significant work is still required to develop opinion diffusion models that accurately reflect social media opinion evolution.

%%%%%%%%%%%%%%%%%%%%%%%%%%%%%%%%%%%%%%%%%%%%%%%%%%%%%%%%%%%%%
% RELATED WORK
%%%%%%%%%%%%%%%%%%%%%%%%%%%%%%%%%%%%%%%%%%%%%%%%%%%%%%%%%%%%%
\section{Related Work}\label{sec:rel_work}
% opinion diffusion models
% high-oder opinion diffusion models
% studies on opinion change
Traditionally, the dynamics of diffusion in online social media—whether related to (mis-)information, innovations, or beliefs—have been studied using graph-based models that abstract pairwise interactions occurring in the online world~\cite{Chamley_MSP_2013,Li_CSUR_2021,Ruffo_CSR_2023,Zhou_CSUR_2021}. Only in recent years has the literature begun to consider group interactions in diffusion models, employing high-order networks such as simplicial complexes and hypergraphs to better capture these dynamics~\cite{Battiston_PhysicsReports_2020}.
Generally, works in this domain extend social and evolutionary diffusion frameworks to such high-order structures. 
%
% committed minorities
For instance, Iacopini et al.~\cite{Iacopini_CommPhys_2022} analyze the role of group interactions on critical mass dynamics, studying how committed minorities of different sizes can overturn apparently stable social norms by extending the naming game framework. 
%
% cooperation
% Civilini_PhysRevLett_2024 possible addition
Alvarez-Rodriguez et al.~\cite{AlvarezRodriguez_NatureHB_2021} et Xu et al.~\cite{Xu_Chaos_2024} focus on group-induced cooperation via evolutionary game approaches, finding that interactions in groups of different sizes affect the evolution of cooperation~\cite{AlvarezRodriguez_NatureHB_2021} and that nested structures of higher-order interactions can promote cooperation~\cite{Xu_Chaos_2024}. 
%
% consensus
Sahasrabuddhe et al.~\cite{Sahasrabuddhe_2021} study the role of group interactions in consensus dynamics and propose a multibody diffusion framework combining homophily and social influence theories (see Section~\ref{subsec:diffusion_model}) to explore the role of involvement and stubbornness on polarization. Papanikolaou et al.~\cite{Papanikolaou_PhysRevE_2022} and Hickok et al.~\cite{Hickok_ADS_2022} present two other representative works related to consensus dynamics by exploiting an adaptive voter model and a bounded confidence model, respectively. Specifically, the former found that group interactions accelerate the formation of consensus, while the latter demonstrate that opinions may \textit{jump} from one opinion cluster to another.

Other works focus on finding key influencer nodes by addressing the influence maximization problem on hypergraphs, where hyperedges model crowd influence or, more generally, shared features among entities of interest. These studies usually generalize the majority rule~\cite{Gangal_IJCAIwrks_2016}, linear threshold~\cite{Qu_IS_2025,Wu_PhysA_2023,Zhu_TCSS_2019}, and epidemic~\cite{Gong_IPM_2024,Suo_PhysicsA_2018} diffusion frameworks. 

%%%%%%%%%%%%%%%%%%%%%%%%%%%%%%%%%%%%%%%%%%%%%%%%%%%%%%%%%%%%%
% CONCLUSION
%%%%%%%%%%%%%%%%%%%%%%%%%%%%%%%%%%%%%%%%%%%%%%%%%%%%%%%%%%%%%
\section{Conclusion}\label{sec:conclusion}
Understanding group-mediated opinion dynamics in social media has become crucial for unraveling offline behaviors, particularly for predicting phenomena like collective actions. While current opinion diffusion models strive to capture the complexity of social interactions, the lack of ground-truth data has limited our ability to evaluate their effectiveness. In this work, we aimed to push a step forward in this direction by providing empirical evidence for the presence and significance of high-order interactions in online discussions. Through our temporal hypergraph model of conversational threads, compared against traditional graph-based approaches, we demonstrate that incorporating high-order social dynamics may lead to more accurate predictions of opinion evolution. Although our findings represent progress in modeling these complex social systems, significant challenges remain in fully capturing the nuances of opinion dynamics in online spaces.

\smallskip
\subsubsection*{Limitations}
% construction validity
% - choise of Reddit and the subreddits
% internal validity
% - the construction of the ground truth
% external validity
% - no real opinion change
%
% construction validity
To ensure the construction validity of our findings, we considered the California wildfires of 2021 and 2022 as a significant exogenous event that generated extensive discussion on Reddit, analyzing the most active subreddits on this topic. While this dataset provides valuable insights into our specific use case, we acknowledge that it may not fully represent the broader spectrum of online communities, potentially limiting the generalizability of our results. We plan to validate our findings across a wider range of online platforms that provide access to user data, such as Bluesky.

% internal validity
\smallskip
To ensure internal validity, we evaluated all three network models under identical conditions, employing the multi-body consensus model developed by Sahasrabuddhe et al.~\cite{Sahasrabuddhe_2021} to simulate opinion dynamics. This model was particularly suitable for our analysis as it can be applied to both hypergraphs and graphs without modification, since graphs can be interpreted as hypergraphs with hyperedges of size two.

The construction of a ground truth presented a significant methodological challenge due to the absence of comprehensive real-world opinion change data. While the SPINOS dataset~\cite{Sakketou_ACL_2022} offers human-validated user opinions about climate change, its fragmentary nature made it unsuitable for constructing fully labeled conversational (hyper)networks. To address this limitation, we developed a synthetic ground truth using GPT-3.5, which represented the state-of-the-art at the time of our experiments. Future work should explore the application of newer large language models for stance detection across diverse topics in online discussions to also enable data-driven studies of opinion dynamics in social media~\cite{pera2025extracting}.

% external validity
\smallskip
Regarding external validity, one last reflection relates to the concept of opinion change. Due to the absence of real-world opinion change data, we cannot assert that the observed opinion drifts in our study necessarily translate to offline behaviors or contexts.

\smallskip
\subsubsection*{Future work}
In future work, we will focus on two key aspects of opinion dynamics in online discussions: the temporal patterns of opinion shifts and cross-community influence. Specifically, we plan to analyze how conversational network structures influence the timing of opinion changes and examine the role of users who participate across multiple subreddits as potential opinion bridges between communities. 

More generally, future research should develop more comprehensive and extensive ground-truth datasets to inform the design, implementation, and evaluation of robust models that can accurately represent real-world opinion dynamics. Further, our results suggest the need to reevaluate the traditional frameworks through which we interpret opinion dynamics models, as the existing approaches rely on assumptions that may not fully align with the complexities revealed by empirical data. This reflection calls for a broader reconsideration of how we conceptualize and model opinion dynamics, particularly in understanding how individual beliefs evolve within complex social systems. Therefore, future research should focus not only on developing more sophisticated modeling techniques but also on challenging our fundamental assumptions about how opinions form and propagate through social media.

\printbibliography



\section*{Supplementary material}

# Modeling the Impact of Group Interactions on Climate-related Opinion Change in Reddit — Supplemental Material —

Alessia Antelmi, Carmine Spegnuolo, and Luca Maria Aiello

## 1 Ground-truth construction: GPT 3.5 Prompt

We used the following prompt to label each comment with the GPT-3.5 Turbo model.

"A sequence of statements (one line, one statement) about climate change as input follows. This topic is about the long-term shift in global or regional climate patterns.

Score each statement with a real value in the continuous range [0,1], where

- The score 0.0 means that the statement is "Strongly against" the stance "Climate change is a real concern.". Against climate change are the people who either believe that it is not a real problem or that it exists but is not man-made, or believe that it exists, is man-made, but it does not threaten (or will ever threaten) human existence.
  - "Strongly against" comments are used to generalize statements and use a provocative tone.
  - "Somewhat against" comments are usually open to discussion.
- The score 1.0 means that the statement is "Strongly in favor" of the stance "Climate change is a real concern.". In favor of climate change are the authors who support that climate change is a real man-made problem that threatens (or will eventually threaten) human existence and affects (or will affect) the survival of animals.
  - "Strongly in favor" comments use assertive and strong language.
  - "Somewhat in favor" comments sound less dogmatic and are willing to engage in conversation.
- The score 0.5 means that the statement expresses a "Neutral" stance toward climate change. Neutral toward climate change are the authors who objectively report facts or news or ask for information, opinion, orientation, and confirmation without revealing their stance on climate change or expressing any subjective opinion.

If the statement does NOT relate to climate change, assign NI as a class."

# 2 Experimental suite: data

Table 1: Number of shared users between different subreddits. The last column reports the number of users evaluated per single subreddit or their combination.

| Subreddits | Unique users | Shared users | Evaluated users |
|---|---|---|---|
| r/AskReddit | 10,953 | - | 2,552 (23,3%) |
| r/bayarea | 1,895 | - | 341 (18%) |
| r/California | 1,743 | - | 260 (14,9%) |
| r/collapse | 21,532 | - | 8,017 (37,2%) |
| r/news | 14,159 | - | 1,832 (12,9%) |
| r/politics | 10,407 | - | 1,414 (13,6%) |
| r/news, r/politics | 22,793 | 1,492 (6.5%) | 3,370 (14,8%) |
| r/bayarea, r/California | 3,433 | 151 (4.4%) | 593 (17,3%) |
| r/collapse, r/news | 34,500 | 966 (2.8%) | 9,863 (28,6%) |
| r/AskReddit, r/news | 24,515 | 465 (1.9%) | 4,436 (18,1%) |
| r/collapse, r/politics | 31,178 | 565 (1.8%) | 9,447 (30,3%) |
| r/AskReddit, r/politics | 20,989 | 259 (1.2%) | 3,997 (19%) |
| r/AskReddit, r/collapse | 32,044 | 365 (1.1%) | 10,582 (33%) |
| r/California, r/news | 15,627 | 172 (1.1%) | 2,111 (13,5%) |
| r/bayarea, r/news | 15,885 | 111 (0.7%) | 2180 (13,7%) |
| r/california, r/collapse | 23,131 | 88 (0.4%) | 8,017 (34,7%) |
| r/collapse, r/news, r/politics | 42,527 | 137 (0.3%) | 11,408 (26,8%) |
| r/askreddit, r/news, r/politics | 32,861 | 70 (0.2%) | 6,003 (18,3%) |
| All subreddits | 54,923 | 40 (0,07%) | 14,673 (26,7%) |



\end{document}